\documentclass{article}
\usepackage{amsmath}

\newtheorem{theorem}{Theorem}[section]
\newtheorem{lemma}[theorem]{Lemma}

\newtheorem{corollary}[theorem]{Corollary}
\newenvironment{proof}[1][Proof]{\begin{trivlist}
\item[\hskip \labelsep {\bfseries #1}]}{\end{trivlist}}

\newenvironment{example}[1][Example]{\begin{trivlist}
\item[\hskip \labelsep {\bfseries #1}]}{\end{trivlist}}
\newtheorem{remark}[theorem]{Remark}
\newcommand{\qed}{\nobreak \ifvmode \relax \else
      \ifdim\lastskip<1.5em \hskip-\lastskip
      \hskip1.5em plus0em minus0.5em \fi \nobreak
      \vrule height0.75em width0.5em depth0.25em\fi}
\numberwithin{equation}{section}

\begin{document}

%\title{The properties of the residual entanglement for n qubits \thanks{%
%The paper was supported by NSFC(Grants No. 60433050 and 60673034) and the
%basic research fund of Tsinghua university No: JC2003043. }}

\centerline{\large\bf An entanglement measure for n
qubits\footnote{The paper was supported by NSFC(Grants No. 60433050
and 60673034) and the basic research fund of Tsinghua university NO:
JC2003043. }} %\thanks{email address:dli@math.tsinghua.edu.cn}
%\footnote{email:dli@math.tsinghua.edu.cn}

\centerline{Dafa Li$^{a}$\footnote{email
address:dli@math.tsinghua.edu.cn}, Xiangrong Li$^{b}$, Hongtao
Huang$^{c}$, Xinxin Li$^{d}$ }

\centerline{$^a$ Dept of mathematical sciences, Tsinghua University,
Beijing 100084 CHINA}

\centerline{$^b$ Department of Mathematics, University of
California, Irvine, CA 92697-3875, USA}

\centerline{$^c$ Electrical Engineering and Computer Science Department} %
\centerline{ University of Michigan, Ann Arbor, MI 48109, USA}

\centerline{$^d$ Dept. of Computer Science, Wayne State
University, Detroit, MI 48202, USA}

%\maketitle

%\maketitle
Pacs: 03.67.Mn, 03.65.Ud.

Abstract

In Phys. Rev. A 61, 052306 (2000), Coffman, Kundu and Wootters\ introduced
the residual entanglement for three qubits.\ In this paper, we present the
entanglement measure $\tau (\psi )$ for even $n$ qubits; for odd $n$ qubits,
we propose the residual entanglement $\tau ^{(i)}(\psi )$\ with respect to
qubit $i$ and the odd $n$-tangle $R(\psi )$ by averaging the residual
entanglement with respect to each qubit. In this paper, we show that these
measures are $LU$-invariant, entanglement monotones, invariant under
permutations of the qubits, and multiplicative in some cases.


%\maketitle
Pacs: 03.67.Mn, 03.65.Ud.

\section{Introduction}

Entanglement plays an important role in quantum computation and quantum
information \cite{Bennett}\cite{Nilsen}. Many researchers in quantum
information theory show interests in entanglement measures.\ Wootters
introduced the idea of concurrence for two qubits to quantify entanglement
\cite{Wootters}. Subsequently, the concurrence was further developed in \cite%
{Uhlmann}\cite{Audenaert}\cite{Mintert}. Recently, Coffman, Kundu, and
Wootters presented the residual entanglement which measures the amount of
entanglement between subsystem $A$ and subsystems $BC$\ for a tripartite
state and gave an elegant expression for computing the residual entanglement
for three qubits via the concurrence \cite{Coffman}. Vidal proposed
entanglement monotone in \cite{Vidal}. It was later proved that the residual
entanglement for three qubits is an entanglement monotone \cite{Dur}.
Recently, many authors have studied the residual entanglement.

Wong and Christensen defined even $n$-tangle for even $n$ qubits which is
invariant under permutations of the qubits and demonstrated that the even $n$%
-tangle for even $n$ qubits is an entanglement monotone \cite{Wong}. Their
even $n$-tangle for even $n$ qubits is listed as follows. See (2) in \cite%
{Wong}.
\begin{equation*}
\tau _{1...n}=2|\sum a_{\alpha _{1}...\alpha _{n}}a_{\beta _{1}...\beta
_{n}}a_{\gamma _{1}...\gamma _{n}}a_{\delta _{1}...\delta _{n}}\times
\epsilon _{\alpha _{1}\beta _{1}}\epsilon _{\alpha _{2}\beta
_{2}}...\epsilon _{\alpha _{n-1}\beta _{n-1}}\epsilon _{\gamma _{1}\delta
_{1}}\epsilon _{\gamma _{2}\delta _{2}}....\times \epsilon _{\gamma
_{n-1}\delta _{n-1}}\epsilon _{\alpha _{n}\gamma _{n}}\epsilon _{\beta
_{n}\delta _{n}}|.
\end{equation*}%
The even $n$-tangle is quartic and requires $3\ast 2^{4n}$ multiplications.
Our entanglement measure $\tau (\psi )$ for even $n$\ qubits is quadratic
and requires $2^{n-1}$ multiplications \cite{LDF07}. Furthermore, Wong and
Christensen indicated that the even $n$-tangle for even $n$ qubits is not a
measure of $n$-way entanglement \cite{Wong}. The n-way entanglement is the
entanglement that critically involves all n particles \cite{Wong}.\ For odd $%
n$ qubits, they said that the $n$-tangle is undefined for odd $n>3$, see
their abstract in \cite{Wong}.

In a separate work \cite{Yu}, Yu and Song defined the residual entanglement
for $n$ qubits as follows.

\begin{equation}
\tau _{ABC...N}=\min \{\tau _{\alpha }|\alpha
=1,2,...,\sum_{i=1}^{[N/2]}C_{N}^{i}\},
\end{equation}%
where $\alpha $ corresponds to all the possible foci and $%
C_{N}^{i}=n!/[(n-i)!i!]$. However, they did not show whether the residual
entanglement is $LU$-invariant, or invariant under permutations of the
qubits. Nor did they show that the residual entanglement is an entanglement
monotone.

In another paper, Osterloh and Siewert constructed an $n$-qubit entanglement
monotone from antilinear operators \cite{Osterloh}.

In an interesting work \cite{Ou2},\ Ou and Fan found that the monogamy of
concurrence implies the monogamy of negativity, and that the resulting
residual entanglement obtained through symmetrization all possible subsystem
permutation gives rise to an entanglement monotone. In \cite{Ou2}, they
defined the negativity $\mathcal{N=(}\left\vert \left\vert \rho
^{T_{A}}\right\vert \right\vert -1)/2$, where $\rho ^{T_{A}}$ is the partial
transpose with respect to the subsystem $A$. Then, \ they defined the
residual entanglements$\ \pi _{A}=$$\mathcal{N}_{A(BC)}^{2}-\mathcal{N}%
_{AB}^{2}-\mathcal{N}_{AC}^{2}$, $\pi _{B}=$$\mathcal{N}_{B(AC)}^{2}-%
\mathcal{N}_{BA}^{2}-\mathcal{N}_{BC}^{2}$, and $\pi _{C}=$$\mathcal{N}%
_{C(AB)}^{2}-\mathcal{N}_{CA}^{2}-\mathcal{N}_{CB}^{2}$. However, they
indicated that the residual entanglement corresponding to the different
focus varies under permutations of the qubits, i.e., generally $\pi _{A}\neq
\pi _{B}\neq \pi _{C}$.

Entanglement monotone is an important quality for entanglement measures. Any
increase in correlations achieved by LOCC should be naturally classical. In
other words, entanglement should be non-increasing under LOCC. Therefore
monotonicity for entanglement measure under LOCC is considered as the
natural requirement \cite{Vidal}. The symmetry of entanglement measure under
permutations implies that the measure represents a collective property of
the qubits which is unchanged by permutations \cite{Coffman}. In this paper,
we present entanglement measures for $n$ qubits, and demonstrate that the
entanglement measures in question are (i) entanglement monotones, i.e.,
non-increasing on average under LOCC in all the $n$ qubits, (ii) invariant
under permutations of the qubits, and (iii) multiplicative in some cases.

In this paper, in Sec. 2 we study the entanglement measure $\tau (\psi )$
for even $n$ qubits. In Sec. 3, we investigate the residual entanglement $%
\tau ^{(i)}(\psi )$\ with respect to qubit $i$ and the odd $n$-tangle $%
R(\psi )$ for odd $n$ qubits. $\tau (\psi )$, $\tau ^{(i)}(\psi )$, and $%
R(\psi )$ only require $+$, $-$, and $\ast $ operations.

Notations: (1). Let $|\psi \rangle $ $=$\ $\sum_{i=0}^{2^{n}-1}a_{i}|i%
\rangle $ and $|\psi ^{\prime }\rangle $ $=$\ $\sum_{i=0}^{2^{n}-1}a_{i}^{%
\prime }|i\rangle $ be states of $n$ qubits in this paper.

\ \ \ \ \ \ \ \ \ (2). Let $i_{n-1}...i_{1}i_{0}$ be an $n-$bit binary
representation of $i$. That is, $i=i_{n-1}2^{n-1}+...+i_{1}2^{1}+i_{0}2^{0}$%
. Then, let $N(i)$ be the number of the occurrences of \textquotedblleft $1$%
\textquotedblright\ in $i_{n-1}...i_{1}i_{0}$\ and $N^{\ast }(i)$ be the
number of the occurrences of \textquotedblleft $1$\textquotedblright\ in $%
i_{n-2}...i_{1}i_{0}$, respectively. \

\section{Entanglement measure for even $n$ qubits}

In our previous work\cite{LDF07}, we defined the entanglement measure of the
state $|\psi \rangle $ of even $n$ qubits as
\begin{equation}
\tau (\psi )=\ 2\left\vert \mathcal{I}^{\ast }(a,n)\right\vert ,
\label{even-residual-def}
\end{equation}
where
\begin{eqnarray}
\mathcal{I}^{\ast }(a,n) & =& \sum_{i=0}^{2^{n-2}-1}sgn^{\ast
}(n,i)(a_{2i}a_{(2^{n}-1)-2i}-a_{2i+1}a_{(2^{n}-2)-2i}).  \label{even-ver-2}
\end{eqnarray}
The functions $sgn$ and $sgn^{\ast }$ have been defined previously in \cite%
{LDF07}. To facilitate reading, we have listed the definitions of $sgn$ and $%
sgn^{\ast }$ in Appendix A. When $n=2$, $\tau (\psi )=2\left\vert
a_{0}a_{3}-a_{1}a_{2}\right\vert $, which is just the concurrence for two
qubits.

Theorem 1 in \cite{LDF07} implies that $\mathcal{I}^{\ast }(a,n)$ and $\tau
(\psi )$ for even $n$\ qubits are invariant under $SL$ (determinant--one)
operators, especially under $LU$ (local unitary) operators. In order to
argue below that $\tau (\psi )$ for even $n$ qubits is an entanglement
monotone, we need the following result. If the states $|\psi ^{\prime
}\rangle $ and $|\psi \rangle $ are related by a local operator as
\begin{equation}
|\psi ^{\prime }\rangle =\underbrace{\alpha \otimes \beta \otimes \gamma
\cdots }_{n}|\psi \rangle ,  \label{even-1}
\end{equation}%
then%
\begin{equation}
\mathcal{I}^{\ast }(a^{\prime },n)=\mathcal{I}^{\ast }(a,n)\underbrace{\det
(\alpha )\det (\beta )\det (\gamma )...}_{n}  \label{even-2}
\end{equation}%
and
\begin{equation}
\tau (\psi ^{\prime })=\tau (\psi )\underbrace{|\det (\alpha )\det (\beta
)\det (\gamma )...|}_{n}.  \label{even-residual-2}
\end{equation}%
\

It is easy to see that Eq. (\ref{even-residual-2}) follows Eqs. (\ref%
{even-residual-def}) and (\ref{even-2}). The proof of Eq. (\ref{even-2}) is
found in part A of Appendix D in \cite{LDF07e} in which the condition that $%
\alpha $, $\beta $, ... are invertible was not used. Following this result,
we have the following two results. (1). That the states $|\psi ^{\prime
}\rangle $ and $|\psi \rangle $ are connected by SLOCC, i.e., $\alpha $, $%
\beta $, ... are invertible, becomes a special case of Eq. (\ref{even-1}).
Hence Eq. (\ref{even-residual-2}) holds. (2). Eq. (\ref{even-residual-2}) is
true even if the states $|\psi ^{\prime }\rangle $ and $|\psi \rangle $ are
connected by general LOCC, i.e., by non-invertible operators (see \cite{Dur}%
).

\subsection{Invariance under permutations of the $n$\ qubits}

For a state of even $n$ qubits, $|\psi \rangle$, we show in this section the
invariance of $\tau(\psi)$ under permutations of the qubits. To this end, we
first prove following propositions.

\bigskip

\begin{remark}
\label{remark1} Each term of $\mathcal{I}^{\ast }(a,n)$ in Eq. (\ref%
{even-ver-2}) takes the form $(-1)^{N(k)}a_{k}a_{2^{n}-1-k}$.
\end{remark}

\begin{proof}
It is easy to see that binary representations of $k$ and $2^{n}-1-k$ are
complementary. So, $N(k)+N(2^{n}-1-k)=n$. Hence, $%
(-1)^{N(k)}=(-1)^{N(2^{n}-1-k)}$. By the definition for $sgn^{\ast }$ in
Appendix A, $sgn^{\ast }(n,i)=(-1)^{N(i)}$ when $0\leq i\leq 2^{n-3}-1$ and $%
sgn^{\ast }(n,i)=(-1)^{n+N(i)}$\ when $2^{n-3}\leq i\leq 2^{n-2}-1$.
Therefore, $sgn^{\ast }(n,i)=(-1)^{N(i)}$ when $n$ is even and $0\leq i\leq
2^{n-2}-1$. Next there are two cases.

\begin{enumerate}
\item Consider term $sgn^{\ast }(n,i)a_{2i}a_{(2^{n}-1)-2i}$. Since$\
N(2i)=N(i)$, this remark is true for case 1.

\item Consider term $-sgn^{\ast }(n,i)a_{2i+1}a_{(2^{n}-2)-2i}$. Since $%
N(2i+1)=N(i)+1$, this remark is true for case 2.
\end{enumerate}
\end{proof}

\bigskip

\begin{lemma}
\label{lemma1} The term $\mathcal{I}^{\ast }(a,n)$ in Eq. (\ref{even-ver-2}%
)\ does not vary under any permutation of the $n$ qubits.
\end{lemma}

\begin{proof}
By remark \ref{remark1}, each term of $\mathcal{I}^{\ast }(a,n)$ is of the
form $(-1)^{N(k)}a_{k}a_{2^{n}-1-k}$. Let the binary number for $k$
correspond to the binary number for $k^{\prime }$ under permutation $\pi $
of the qubits. Then, the binary number for $2^{n}-1-k$ corresponds to the
binary number for $2^{n}-1-k^{\prime }$ under $\pi $. That is, $\ \pi
(2^{n}-1-k)=2^{n}-1-k^{\prime }$. Obviously, $a_{k}=a_{k^{\prime }}$, $%
a_{2^{n}-1-k}=a_{2^{n}-1-k^{\prime }}$, and $N(k)=N(k^{\prime })$. Thus, $%
(-1)^{N(k)}a_{k}a_{2^{n}-1-k}=(-1)^{N(k^{\prime })}a_{k^{\prime
}}a_{2^{n}-1-k^{\prime }}$. Therefore, $\mathcal{I}^{\ast }(a,n)$ does not
vary under any permutation of the qubits.
\end{proof}

\noindent From lemma \ref{lemma1} and Eq. (\ref{even-residual-def}), we have
the following corollary 1.

\medskip

\begin{corollary}
\label{corollary1} The residual entanglement $\tau (\psi )$ does not vary
under any permutation of the $n$ qubits.
\end{corollary}

\subsection{Product states}

For product states, the residual entanglement $\tau (\psi )$ either
vanishes or is multiplicative. In this section, we state an
important theorem and refer the reader to the Appendix B for a
detailed proof.

\bigskip

\begin{theorem}
\label{theorem1} Let $|\psi \rangle $ be a state of even $n$ qubits which
can be expressed as a tensor product state of state $|\phi \rangle $ of the
first $l$ qubits and state $|\omega \rangle $ of the rest $(n-l)$ qubits.
Let $|\phi \rangle $ $=$\ $\sum_{i=0}^{2^{l}-1}b_{i}|i\rangle $, where $%
1\leq l<n$, and $|\omega \rangle $ $=$\ $\sum_{i=0}^{2^{n-l}-1}c_{i}|i%
\rangle $. Then, $\tau (\psi )=\tau (\phi )\tau (\omega )$ for even $l$
while $\tau (\psi )=0$ for odd $l$.
\end{theorem}

\begin{proof}
See Appendix B for a detailed proof.
\end{proof}

It is instructive to look at several examples to see the usefulness of this
theorem. In example 1, we show a four-qubit state in which $\tau (\psi )=1$
and in example 2, we look at a case of a six-qubit state in which $\tau
(\psi )=0$.

\begin{example}
\label{ex1} For four qubits, $\tau ((1/2)((|00\rangle +|11\rangle
)_{12}\otimes (|00\rangle +|11\rangle )_{34}))=1$.
\end{example}

\begin{example}
\label{ex2} For six qubits, $\tau ((1/2)((|000\rangle +|111\rangle
)_{123}\otimes (|000\rangle +|111\rangle )_{456}))=0$.
\end{example}

\noindent It is possible to extend theorem 1 further. From theorem 1 and
corollary 1, we have the following corollary \ref{corollary2}:

\begin{corollary}
\label{corollary2} (An extension of theorem 1): (1). If $|\psi \rangle $ is
a tensor product state of state $|\phi \rangle $ of even qubits and state $%
|\omega \rangle $ of even qubits, then $\tau (\psi )=\tau (\phi )\tau
(\omega )$. That is, $\tau (\psi )$ is multiplicative. (2). If $|\psi
\rangle $ is a tensor product state of state $|\phi \rangle $ of odd qubits
and state $|\omega \rangle $ of odd qubits, then $\tau (\psi )=0$.
\end{corollary}

The corollary \ref{corollary2} argues that $\tau (\psi )$ for even $n$
qubits is not a measure of $n$-way entanglement. Note that the conjecture
for even $n$ qubits in \cite{LDF07} is the same as Corollary 2. At this
juncture, it is probably interesting to note some examples for six-qubit
states.

\begin{example}
\label{ex3} For six qubits, $\tau ((1/2)((|0000\rangle +|1111\rangle
)_{1456}\otimes (|00\rangle +|11\rangle )_{23}))=1$.
\end{example}

\begin{example}
\label{ex4} For six qubits, $\tau ((1/2)((|000\rangle +|111\rangle
)_{135}\otimes (|000\rangle +|111\rangle )_{246}))=0$.
\end{example}

In \cite{Dur} SLOCC classes of three qubits are related by means of
non-invertible operators, i.e., of general LOCC, see Fig.1 in \cite{Dur}.
Unfortunately, we can not derive a nice result for four qubits. For example,
for four qubits, no non-invertible operators can transform the state $%
|GHZ\rangle $ to a state within $|GHZ\rangle _{12}\otimes |GHZ\rangle _{34}$
SLOCC class. Assume that the states $|\phi \rangle $ and $|GHZ\rangle $ are
connected by a non-invertible operator as $|\phi \rangle =\alpha \otimes
\beta \otimes \gamma \otimes \delta $ $|GHZ\rangle $. Then by Eq. (\ref%
{even-residual-2}), $\tau (\phi )=\tau (GHZ)|\det (\alpha )\det (\beta )\det
(\gamma )\det (\delta )|=0$. However, for any state $|\phi \rangle $ in $%
|GHZ\rangle _{12}\otimes |GHZ\rangle _{34}$ SLOCC\ class, $\tau (\phi )\neq
0 $ by Eq. (\ref{even-residual-2}) and Example 1.

\subsection{Entanglement monotone}

As indicated in \cite{Vidal}, a natural measure of entanglement should also
be an entanglement monotone. Let us follow the idea in \cite{Dur} to prove
that $\tau (\psi )$ for $n$ qubits is an entanglement monotone. Based on the
work in \cite{Dur}, it is enough to consider two-outcome POVM's and apply
POVM's to one party. For example, we can simply apply a local POVM to qubit $%
k$. Let $A_{1}$ and $A_{2}$ be the two POVM elements such that $%
A_{1}^{+}A_{1}+A_{2}^{+}A_{2}=I$. By the singular value decomposition, there
are unitary matrices $U_{i}$ and $V_{i}$ and diagonal matrices $D_{i}$ with
non-negative entries such that $A_{i}=U_{i}D_{i}V_{i}$ \cite{Nilsen}, where $%
D_{1}=diag(a,b)$ and $D_{2}=diag((1-a^{2})^{1/2},(1-b^{2})^{1/2})$ \cite{Dur}%
. Let $|\psi \rangle $ be an initial state and
\begin{equation}
|\bar{\phi}_{i}\rangle =\underbrace{I\otimes ...\otimes I}_{k-1}\otimes
A_{i}\otimes \underbrace{I\otimes ...\otimes I}_{n-k}|\psi \rangle
\label{POVM1}
\end{equation}%
be the states after the application of the POVM for any $n$ qubits, where $I$
is an identity. To normalize $|\bar{\phi}_{i}\rangle $, let $|\phi
_{i}\rangle =|\bar{\phi}_{i}\rangle /\sqrt{p_{i}}$, where $p_{i}=\langle
\bar{\phi}_{i}|\bar{\phi}_{i}\rangle $. Clearly $p_{1}+p_{2}=1$ \cite{Nilsen}%
. As discussed in \cite{Dur}, next we can consider
\begin{equation}
\langle \tau ^{\eta }\rangle =p_{1}\tau ^{\eta }(\phi _{1})+p_{2}\tau ^{\eta
}(\phi _{2})\text{, where }0<\eta \leq 1\text{,}  \label{Monotone1}
\end{equation}%
and prove
\begin{equation}
\langle \tau ^{\eta }\rangle \leq \tau ^{\eta }(\psi )
\end{equation}%
to show that $\tau $ is an entanglement monotone.

It is intuitive that $\tau (\phi _{i})=\tau (\bar{\phi}_{i})/p_{i}$ because $%
\tau $ is a quadratic function with respect to its coefficients in the
standard basis, see Eq. (\ref{even-ver-2}). Note that $\tau $ is a quartic
function in \cite{Dur}\cite{Wong}. By Eqs. (\ref{even-residual-2}) and (\ref%
{POVM1}),
\begin{equation}
\tau (\bar{\phi}_{i})=\tau (\psi )\left\vert \det (A_{i})\right\vert =\tau
(\psi )\left\vert \det (D_{i})\right\vert .
\end{equation}%
So, it is trivial to get $\tau (\bar{\phi}_{1})=ab\tau (\psi )$ and $\tau (%
\bar{\phi}_{2})=[(1-a^{2})(1-b^{2})]^{1/2}\tau (\psi )$. By substituting $%
\tau (\bar{\phi}_{1})$\ and $\tau (\bar{\phi}_{2})$ into Eq. (\ref{Monotone1}%
), we get
\begin{equation}
\langle \tau ^{\eta }\rangle =\{p_{1}\frac{(ab)^{\eta }}{p_{1}^{\eta }}+p_{2}%
\frac{[(1-a^{2})(1-b^{2})]^{\eta /2}}{p_{2}^{\eta }}\}\tau ^{\eta }(\psi ).
\end{equation}%
When $\eta =1$,
\begin{equation}
\langle \tau \rangle =\{ab+[(1-a^{2})(1-b^{2})]^{1/2}\}\tau (\psi ).
\end{equation}%
As discussed in \cite{Dur}, it is easy to derive $\langle \tau \rangle \leq
\tau (\psi )$. Thus, this means when $\eta =1$, $\tau $ is an entanglement
monotone. Finally, as pointed out in \cite{Dur}, when $0<\eta \leq 1$, it is
easy to show that $\tau $ is an entanglement monotone.

It is worthwhile pointing out that in \cite{Dur} the authors simplified the
calculation for $\tau (\bar{\phi}_{i})$ \cite{Dur}\cite{Wong} by using the
restriction $V_{1}=V_{2}$ since they apparently thought the fact that $A_{1}$
and $A_{2}$ constitute a POVM implies $V_{1}=V_{2}$. The authors in \cite%
{Dur}\ and \cite{Wong}\ used the invariance of the 3-tangle in \cite{Dur}
and the even $n$-tangle in \cite{Wong} under permutations of the qubits,
respectively, to consider a local POVM in party A only. Moreover, they also
used the invariance of the 3-tangle and the even $n$-tangle under $LU$
respectively to obtain $\tau (U_{i}D_{i}V\psi )=\tau (D_{i}V\psi )$ in \cite%
{Dur}\cite{Wong}.

\section{Residual entanglement for odd $n$ qubits and the odd $n$-tangle}

In this section, we propose the residual entanglement with respect to each
qubit. We consider $\tau (\psi )$ for odd $n$ qubits in \cite{LDF07} as the
residual entanglement with respect to qubit 1. Let $(1,i)$ be the
transposition of qubits $1$ and $i$, and $(1,i)|\psi \rangle $ be the state
obtained from $|\psi \rangle $ under the transposition $(1,i)$. Let $\tau
^{(i)}(\psi )=\tau ((1,i)\psi )$, $i=2$, $3$, $...$, $n$ and $\tau
^{(1)}(\psi )=\tau (\psi )$. Then, we propose $\tau ^{(i)}(\psi )$\ as the
residual entanglement with respect to qubit $i$, where $i=1,...,n$. It seems
that the residual entanglement $\tau ^{(1)}(\psi )$ with respect to qubit 1
is transferred to qubit $i$ under the transposition $(1,i)$. By averaging
the residual entanglement with respect to each qubit,\ we propose the
following $R(\psi )$ as the odd $n$-tangle.

\begin{equation}
R(\psi )=\frac{1}{n}\sum_{i=1}^{n}\tau ^{(i)}(\psi ).  \label{odd-n-tangle}
\end{equation}%
First, we study the properties of $\tau (\psi )$. Then, by means of the
properties of $\tau (\psi )$, we investigate the residual entanglement $\tau
^{(i)}(\psi )$ with respect to qubit $i$ and the odd $n$-tangle $R(\psi )$.

In \cite{LDF07}, we defined the entanglement measure for the state $|\psi
\rangle $ of odd $n$ qubits as
\begin{equation}
\tau (\psi )=4|(\overline{\mathcal{I}}(a,n))^{2}-4\mathcal{I}^{\ast }(a,n-1)%
\mathcal{I}_{+2^{n-1}}^{\ast }(a,n-1)|,  \label{odd-residual-def}
\end{equation}%
where
\begin{eqnarray}
\overline{\mathcal{I}}(a,n)& = &
\sum_{i=0}^{2^{n-3}-1}sgn(n,i)[(a_{2i}a_{(2^{n}-1)-2i}-a_{2i+1}a_{(2^{n}-2)-2i})
\notag \\
&&-(a_{(2^{n-1}-2)-2i}a_{(2^{n-1}+1)+2i}-a_{(2^{n-1}-1)-2i}a_{2^{n-1}+2i})],
\label{odd-iv}
\end{eqnarray}

\begin{eqnarray}  \label{odd-invariant-2}
&&\mathcal{I}_{+2^{n-1}}^{\ast }(a,n-1)=\sum_{i=0}^{2^{n-3}-1}sgn^{\ast
}(n-1,i)\times  \notag \\
&&(a_{2^{n-1}+2i}a_{(2^{n}-1)-2i}-a_{2^{n-1}+1+2i}a_{(2^{n}-2)-2i}),  \notag
\\
\end{eqnarray}

\
\begin{eqnarray}  \label{odd-invariant-3}
&&\mathcal{I}^{\ast }(a,n-1)= \sum_{i=0}^{2^{n-3}-1}sgn^{\ast }(n-1,i)\times
\notag \\
&&(a_{2i}a_{(2^{n-1}-1)-2i}-a_{2i+1}a_{(2^{n-1}-2)-2i}).  \notag \\
\end{eqnarray}

For $n=3$, $\tau (\psi )$ in Eq. (\ref{odd-residual-def})\ is just simply
the residual entanglement for three qubits \cite{Dur}, i.e., 3 tangle, which
is $\tau _{ABC}=4\left\vert d_{1}-2d_{2}+4d_{3}\right\vert $, where the
expressions for $d_{i}$ are omitted here.

Theorem 2 in \cite{LDF07} implies that $(\overline{\mathcal{I}}(a,n))^{2}-4%
\mathcal{I}^{\ast }(a,n-1)\mathcal{I}_{+2^{n-1}}^{\ast }(a,n-1)$ and $\tau
(\psi )$ are invariant under $SL$-operators, especially under $LU$%
-operators. We argue below that the entanglement measure $\tau $ for odd $n$
qubits is an entanglement monotone, using the following result.

If the states $|\psi ^{\prime }\rangle $ and $|\psi \rangle $ are connected
by a local operator as
\begin{equation}
|\psi ^{\prime }\rangle =\underbrace{\alpha \otimes \beta \otimes \gamma
\cdots }_{n}|\psi \rangle ,  \label{odd-residual-1}
\end{equation}%
then%
\begin{eqnarray}
&&(\overline{IV}(a^{\prime },n))^{2}-4IV^{\ast }(a^{\prime
},n-1)IV_{+2^{n-1}}^{\ast }(a^{\prime },n-1)=  \notag \\
&&[(\overline{IV}(a,n))^{2}-4IV^{\ast }(a,n-1)IV_{+2^{n-1}}^{\ast
}(a,n-1)]\times  \notag \\
&&\underbrace{(\det (\alpha )\det (\beta )\det (\gamma )...)^{2}}_{n},
\label{oddIV3}
\end{eqnarray}%
and
\begin{equation}
\tau (\psi ^{\prime })=\tau (\psi )\underbrace{|\det (\alpha )\det (\beta
)\det (\gamma )...|^{2}}_{n}.  \label{odd-residual-2}
\end{equation}

It is easy to know that Eq. (\ref{odd-residual-2}) follows Eqs. (\ref%
{odd-residual-def}) and (\ref{oddIV3}). For the proof of Eq. (\ref{oddIV3}),
see the proof in part B of Appendix D in \cite{LDF07e} in which the
condition that $\alpha $, $\beta $, ... are invertible was not used.
Following this result, for odd $n$ qubits we also have the following two
results. (1). That the states $|\psi ^{\prime }\rangle $ and $|\psi \rangle $
are connected by SLOCC, i.e., $\alpha $, $\beta $, ... are invertible,
becomes a special case of Eq. (\ref{odd-residual-1}). Hence Eq. (\ref%
{odd-residual-2}) holds. (2). Eq. (\ref{odd-residual-2}) is true even if the
states $|\psi ^{\prime }\rangle $ and $|\psi \rangle $ are connected by
general LOCC, i.e., by non-invertible operators (see \cite{Dur}).

\subsection{Invariance of $\protect\tau (\protect\psi )$ under any
permutation of the qubits $2,3,...,$ $n$.}

The residual entanglement $\tau (\psi )$ is invariant under permutation of
these qubits. To prove the invariance, we prove the following remark \ref%
{remark2} and lemma \ref{lemma2}, and corollary \ref{corollary3} stated
below.

\bigskip

\begin{remark}
\label{remark2} Let $|\psi \rangle $ be a state of odd $n$ qubits. Then each
term of $\overline{\mathcal{I}}(a,n)$ in Eq. (\ref{odd-iv}) is of the form $%
(-1)^{N^{\ast }(k)}a_{k}a_{2^{n}-1-k}$.
\end{remark}

\begin{proof}
Since the binary representations of $k$ and $2^{n}-1-k$ are complementary, $%
N(k)+N(2^{n}-1-k)=n$ and $N^{\ast }(k)+N^{\ast }(2^{n}-1-k)=n-1$. Hence, $%
(-1)^{N^{\ast }(k)}=(-1)^{N^{\ast }(2^{n}-1-k)}$. \ Note that $%
sgn(n,i)=(-1)^{N(i)}$ when $0\leq i\leq 2^{n-3}-1$ by the definition for $%
sgn^{\ast }$ in Appendix A. Next there are four cases.

\medskip

\begin{enumerate}
\item Term $sgn(n,i)a_{2i}a_{(2^{n}-1)-2i}$. Since $0\leq 2i\leq 2^{n-2}-2$,
$N^{\ast }(2i)=N(2i)=N(i)$.

\item Term $-sgn(n,i)a_{2i+1}a_{(2^{n}-2)-2i}$. Since $1\leq 2i+1\leq
2^{n-2}-1$, $N^{\ast }(2i+1)=N(2i+1)=N(i)+1$.

\item Term $-sgn(n,i)a_{(2^{n-1}-2)-2i}a_{(2^{n-1}+1)+2i}$. Clearly, $%
N^{\ast }(2^{n-1}+1+2i)=N^{\ast }(1+2i)=N(i)+1$.

\item Term $sgn(n,i)a_{(2^{n-1}-1)-2i}a_{2^{n-1}+2i}$. \ It is trivial that $%
N^{\ast }(2^{n-1}+2i)=N(2i)=N(i)$.
\end{enumerate}

\noindent Since the above four cases exhaust all possibilities, the remark
holds.
\end{proof}

\bigskip

\begin{lemma}
\label{lemma2} Let $|\psi \rangle $ be a state of odd $n$ qubits. Then, $%
\overline{\mathcal{I}}(a,n)$ in Eq. (\ref{odd-iv})\ does not vary under any
permutation of the qubits 2, 3, ... , and $n$.
\end{lemma}

\begin{proof}
By remark \ref{remark2}, each term of $\overline{\mathcal{I}}(a,n)$ in Eq. (%
\ref{odd-iv}) is of the form $(-1)^{N^{\ast }(k)}a_{k}a_{2^{n}-1-k}$. Let
the binary number for $k$ correspond to the binary number for $k^{\prime }$
under permutation $\pi $ of the qubits 2, 3, ... , and $n$. Then, the binary
number for $2^{n}-1-k$ corresponds to the binary number for $%
2^{n}-1-k^{\prime }$ under $\pi $. That is, $\pi
(2^{n}-1-k)=2^{n}-1-k^{\prime }$. Obviously, $a_{k}=a_{k^{\prime }}$, $%
a_{2^{n}-1-k}=a_{2^{n}-1-k^{\prime }}$, and $N^{\ast }(k)=N^{\ast
}(k^{\prime })$. Thus, $(-1)^{N^{\ast }(k)}a_{k}a_{2^{n}-1-k}=(-1)^{N^{\ast
}(k^{\prime })}a_{k^{\prime }}a_{2^{n}-1-k^{\prime }}$. Therefore, $%
\overline{\mathcal{I}}(a,n)$ does not vary under any permutation of the
qubits 2, 3, ... , and $n$.
\end{proof}

Finally, we have the following corollary concerning the invariance of the
entanglement measure $\tau (\psi )$ under any permutations of the qubits 2,
3, ... , and $n$.

\bigskip

\begin{corollary}
\label{corollary3} Let $|\psi \rangle $ be a state of odd $n$ qubits. Then, $%
\tau (\psi )$ does not vary under any permutation of the qubits 2, 3, ... ,
and $n$.
\end{corollary}

\begin{proof}
Note that a binary representation of each subscript in each term of $%
\mathcal{I}^{\ast }(a,n-1)$ in Eq. (\ref{odd-invariant-3}) is of the form $%
0k_{n-2}...k_{1}k_{0}$ and a binary representation of each subscript in each
term of $\mathcal{I}_{+2^{n-1}}^{\ast }(a,n-1)$ in Eq. (\ref{odd-invariant-2}%
)\ is of the form $1k_{n-2}...k_{1}k_{0}$.\ Under any permutation of the
qubits 2, 3, ... , and $n$,\ by lemma \ref{lemma1} either $\mathcal{I}^{\ast
}(a,n-1)$ or $\mathcal{I}_{+2^{n-1}}^{\ast }(a,n-1)$ does not vary and by
lemma \ref{lemma2} $\overline{\mathcal{I}}(a,n)$ does not vary. Hence, by
the definition in Eq. (\ref{odd-residual-def}),\ $\tau (\psi )$ does not
vary under any permutation of the qubits 2, 3, ... , and $n$.
\end{proof}

\noindent To see the usefulness of the results that we have shown, it is
instructive to study an example:

\begin{example}
\label{example5} Let $|\psi \rangle =(1/2)((|00\rangle +|11\rangle
)_{12}\otimes (|000\rangle +|111\rangle )_{345})$. Then, by Eq. (\ref%
{odd-residual-def}), a simple calculation shows that $\tau (\psi )=0$. Under
the transposition $(1,5)$\ of the qubits 1 and 5, $|\psi \rangle $ becomes $%
|\psi ^{\prime }\rangle =(1/2)((|00\rangle +|11\rangle )_{25}\otimes
(|000\rangle +|111\rangle )_{134})$. By Eq. (\ref{odd-residual-def}), $\tau
(\psi ^{\prime })=1$.
\end{example}

\subsection{Product states}

For product states, $\tau (\psi )$ vanishes or is multiplicative. To prove
this statement, we have the following theorem:

\begin{theorem}
\label{theorem2}Let $|\psi \rangle $ be a state of odd $n$ qubits and a
tensor product state of the state $|\phi \rangle $ of the first $l$ qubits
and the state $|\omega \rangle $ of the rest $(n-l)$ qubits. Let $|\phi
\rangle $ $=$\ $\sum_{i=0}^{2^{l}-1}b_{i}|i\rangle $, where $1\leq l<n$, and
$|\omega \rangle $ $=$\ $\sum_{i=0}^{2^{n-l}-1}c_{i}|i\rangle $. Then, $\tau
(\psi )=\tau (\phi )\tau ^{2}(\omega )$ for odd $l$, while $\tau (\psi )=0$
for even $l$.\
\end{theorem}

\begin{proof}
See Appendix C for the detailed proof.
\end{proof}

\noindent It is interesting to study some examples to see some application
of the theorem.

\begin{example}
\label{ex6} For five qubits, $\tau ((1/2)((|000\rangle +|111\rangle
)_{123}\otimes (|00\rangle +|11\rangle )_{45}))=1$.
\end{example}

\begin{example}
\label{ex7} For five qubits, $\tau ((1/2)((|00\rangle +|11\rangle
)_{12}\otimes (|000\rangle +|111\rangle )_{345}))=0$.
\end{example}

Moreover, from theorem \ref{theorem2} and corollary \ref{corollary3}, we
have the following corollary as an extension of theorem \ref{theorem2}.

\begin{corollary}
\label{corollary4} Theorem \ref{theorem2} holds under any permutation $\pi $
of the qubits $2$, $3$, ... , and $n$. That is, let $|\phi \rangle $ be a
state of $l$ qubits including qubit 1 and state $|\omega \rangle $ be a
state of the rest $(n-l)$ qubits, then, $\tau (\psi )=\tau (\phi )\tau
^{2}(\omega )$ for odd $l$ while $\tau (\psi )=0$ for even $l$. Hence, $\tau
(\psi )$ can be considered to be multiplicative for odd $l$.
\end{corollary}

The corollary \ref{corollary4} implies that for odd $n$ qubits, $\tau (\psi
) $ is not a measure of $n$-way entanglement. In \cite{LDF07}, we
conjectured that $\tau (\psi )=0$ whenever $\psi $ is a product states of
the odd $n$ qubits. This corollary indicates that the conjecture is not
always true.

\begin{example}
\label{ex8} For five qubits, $\tau ((1/2)((|000\rangle +|111\rangle
)_{125}\otimes (|00\rangle +|11\rangle )_{34}))=1$ and $\tau
((1/2)((|00\rangle +|11\rangle )_{15}\otimes (|000\rangle +|111\rangle
)_{234}))=0$.
\end{example}

For five qubits, by resorting to the iterative formula about the number of
the degenerate SLOCC\ classes in \cite{LDF07a}, there are $5\times t(4)+66$
degenerate SLOCC\ classes, where $t(4)$ is the number of true SLOCC
entanglement classes for four qubits. In \cite{LDF07a}, $28$ true SLOCC\
classes for four qubits were found. Hence, in total, there are at least 206
degenerate SLOCC\ classes for five qubits. Note that degenerate SLOCC\
classes are SLOCC\ classes of product states.

By corollary \ref{corollary4}, for five qubits, $\tau $ always vanishes for
all the product states except for the states within the following SLOCC\
classes:

$(|000\rangle +|111\rangle )_{123}\otimes (|00\rangle +|11\rangle )_{45}$, $%
(|000\rangle +|111\rangle )_{124}\otimes (|00\rangle +|11\rangle )_{35}$,

$(|000\rangle +|111\rangle )_{125}\otimes (|00\rangle +|11\rangle )_{34}$, $%
(|000\rangle +|111\rangle )_{134}\otimes (|00\rangle +|11\rangle )_{25}$,

$(|000\rangle +|111\rangle )_{135}\otimes (|00\rangle +|11\rangle )_{24}$, $%
(|000\rangle +|111\rangle )_{145}\otimes (|00\rangle +|11\rangle )_{23}$.

As discussed in \cite{Dur}, SLOCC classes of three qubits are related by
means of non-invertible operators, i.e., of general LOCC, see Fig.1 in \cite%
{Dur}. Here, we want to show that it is not true for five qubits. For
example, no non-invertible operators can transform the state $|GHZ\rangle $
to a state within $|GHZ\rangle _{123}\otimes |GHZ\rangle _{45}$ SLOCC class.
Assume that the states $|\phi \rangle $ and $|GHZ\rangle $ are connected by
a non-invertible operator as $|\phi \rangle =\alpha \otimes \beta \otimes
\gamma \otimes \delta \otimes \eta $ $|GHZ\rangle $. Then by Eq. (\ref%
{odd-residual-2}), $\tau (\phi )=\tau (\psi )|\det^{2}(\alpha
)\det^{2}(\beta )\det^{2}(\gamma )\det^{2}(\delta )\det^{2}(\eta )|=0$.
However, for any state $|\phi \rangle $ in $|GHZ\rangle _{123}\otimes
|GHZ\rangle _{45}$ SLOCC class, $\tau (\phi )\neq 0$ \cite{LDF07}.

\subsection{Entanglement monotone.}

It is easy to see that the first paragraph of Sec. 2.3 is true for any $n$
qubits. It is not hard to show that $\tau (\phi _{i})=\tau (\bar{\phi}%
_{i})/p_{i}^{2}$ because $\tau $ is a quartic function with respect to its
coefficients in the standard basis, see Eqs. (\ref{odd-residual-def})-(\ref%
{odd-invariant-3}). By Eqs. (\ref{POVM1}) and (\ref{odd-residual-2}),
\begin{equation}
\tau (\bar{\phi}_{i})=\tau (\psi )|\det (A_{i})|^{2}=\tau (\psi )\left\vert
\det (D_{i})\right\vert ^{2}.
\end{equation}%
So, $\tau (\bar{\phi}_{1})=(ab)^{2}\tau (\psi )$ and $\tau (\bar{\phi}%
_{2})=(1-a^{2})(1-b^{2})\tau (\psi )$. By substituting $\tau (\bar{\phi}%
_{1}) $\ and $\tau (\bar{\phi}_{2})$ into Eq. (\ref{Monotone1}), we get
\begin{equation}
\langle \tau ^{\eta }\rangle =\{p_{1}\frac{(ab)^{2\eta }}{p_{1}^{2\eta }}%
+p_{2}\frac{[(1-a^{2})(1-b^{2})]^{\eta }}{p_{2}^{2\eta }}\}\tau ^{\eta
}(\psi ).  \label{odd-monotone}
\end{equation}%
Eq. (\ref{odd-monotone}) was also obtained in \cite{Dur}. Therefore the rest
of the proof is the same as the one in \cite{Dur}.

Note that in the above proof, we do not use the restriction $V_{1}=V_{2}$,
the invariance of $\tau $ under permutations of the qubits, or the
invariance of $\tau $\ under $LU$. Therefore, it is not necessary to
establish a relation between the invariance of a measure under permutations
of the qubits and an entanglement monotone.

\subsection{The residual entanglement with respect to each qubit and the odd
$n$-tangle}

\subsubsection{The residual entanglement $\protect\tau ^{(i)}(\protect\psi )$%
\ with respect to qubit $i$}

It is plain to derive that $\tau ^{(i)}(\psi )$ satisfy Eq. (\ref%
{odd-residual-2}). From the properties of $\tau (\psi )$, one can obtain
that (1). $0\leq \tau ^{(i)}(\psi )\leq 1$; (2). $\tau ^{(i)}(\psi )$ are $%
SL $-invariant, especially $LU$-invariant; (3). $\tau ^{(i)}(\psi )$ are
entanglement monotones. (4). $\tau ^{(i)}(\psi )$, $i=1$, $2$, ... , $n$,
are invariant under permutations of the qubits: $1$, ..., $(i-1)$, $(i+1)$,
..., $n$; (5). When $|\psi \rangle $ is a product state of odd $n$ qubits,
that is, $|\psi \rangle =|\phi \rangle \otimes |\omega \rangle $, where $%
|\phi \rangle $ is a state of $l$ qubits including qubit $i$, and $|\omega
\rangle $ is a state of $(n-l)$ qubits, then $\tau ^{(i)}(\psi )=\tau
^{(i)}(\phi )(\tau ^{(i)}(\omega ))^{2}$ for odd $l$ while $\tau ^{(i)}(\psi
)=0$ for even $l$.

We argue that the above (5) holds as follows. Let $(1,i)$ be a transposition
of qubits $1$ and $i$, and the state $(1,i)|\psi \rangle $ be obtained from $%
|\psi \rangle $ under the transposition $(1,i)$. It is not hard to see that $%
\tau ^{(i)}(\psi )=\tau ((1,i)|\psi \rangle )=\tau (((1,i)|\phi \rangle
\otimes (1,i)|\omega \rangle ))$. There are two cases. Case 1. In this case,
qubit 1\ occurs in $|\phi \rangle $. Under the transposition $(1,i)$, qubits
$1$ and $i$ occur in $(1,i)|\phi \rangle $, and $(1,i)|\omega \rangle
=|\omega \rangle $. Case 2. In this case, qubit $1$\ occurs in $|\omega
\rangle $. Under the transposition $(1,i)$, qubit 1 occurs in $(1,i)|\phi
\rangle $ while qubit $i$ occurs in $(1,i)|\omega \rangle $. In either case,
by corollary 4, $\tau (((1,i)|\phi \rangle \otimes (1,i)|\omega \rangle ))=$
$\tau ((1,i)|\phi \rangle )\tau ^{2}((1,i)|\omega \rangle )=$ $\tau
^{(i)}(\phi )(\tau ^{(i)}(\omega ))^{2}$ for odd $l$ while $\tau
(((1,i)|\phi \rangle \otimes (1,i)|\omega \rangle ))=0$ for even $l$. For
the proofs of (1), (2), (3), and (4), see \cite{LDF08}.

\subsubsection{The odd $n$-tangle}

It is not difficult to show that $R(\psi )$ in \ Eq. (\ref{odd-n-tangle})
satisfies Eq. (\ref{odd-residual-2}). Thus, from the properties of $\tau
^{(i)}(\psi )$,\ one can derive that (1). $0\leq R\leq 1$; (2). $R$ is
invariant under $SL$-operators, especially $LU$-operators; (3). $R$ is an
entanglement monotone; (4). $R(\psi )$ is invariant under any permutation of
all the odd $n$\ qubits. For the proofs of (1), (2), (3), and (4), see \cite%
{LDF08}. However, $R(\psi )$ is not multiplicative.

Next let us see the performance of $R(\psi )$ for three qubits. Let $n=3$.
As discussed before, $\tau (\psi )$ happens to be Coffman et al.'s residual
entanglement for three qubits. From (5) of p. 429 in \cite{LDF06}, $\tau
(\psi )=\tau ^{(1)}(\psi )=\tau ^{(2)}(\psi )=\tau ^{(3)}(\psi )$. Thus, $%
R(\psi )=$ $\tau (\psi )$. That is, $R(\psi )$ is just Coffman et al.'s
residual entanglement for three qubits.\

\section{Summary}

We summarize this paper as follows. We demonstrate that the entanglement
measure $\tau (\psi )$ for even $n$ qubits, the residual entanglement $\tau
^{(i)}(\psi )$\ with respect to qubit $i$ and the odd $n$-tangle $R(\psi )$
for odd $n$ qubits satisfy the following properties. (1). $\tau (\psi )$, $%
\tau ^{(i)}(\psi )$, and $R(\psi )$ are between $0$ and $1$; (2). $\tau
(\psi )$, $\tau ^{(i)}(\psi )$, and $R(\psi )$ are $SL$-invariant,
especially $LU$-invariant; (3). $\tau (\psi )$, $\tau ^{(i)}(\psi )$, and $%
R(\psi )$ are entanglement monotones; (4). $\tau (\psi )$ for even $n$
qubits and the odd $n$-tangle $R(\psi )$ are invariant under permutations of
all the qubits; however $\tau ^{(i)}(\psi )$ are invariant only under
permutations of the qubits: $1$,..., $(i-1)$, $(i+1)$, ..., $n$. (5). For
product states, i.e., $|\psi \rangle =|\phi \rangle \otimes |\omega \rangle $%
, for even $n$ qubits, if $|\phi \rangle $ is a state of even qubits then $%
\tau (\psi )=\tau (\phi )\tau (\omega )$ else $\tau (\psi )=0$; for odd $n$
qubits, if $|\phi \rangle $ is a state of $l$ qubits including qubit $i$,
then $\tau ^{(i)}(\psi )=\tau ^{(i)}(\phi )(\tau ^{(i)}(\omega ))^{2}$ for
odd $l$ while $\tau (\psi )=0$ for even $l$.

Monotonicity is a natural requirement for entanglement measure. The symmetry
of entanglement measure under permutations represents a collective property
of the qubits. Therefore the entanglement measures presented in this paper
are natural.

\textbf{Acknowledgement }

Thank L. C. Kwek for his assistance with the language and the formatting of
this manuscript.

\section*{Appendix A. Properties of $sgn$ and $sgn^{\ast }$}

\setcounter{equation}{0} \renewcommand{\theequation}{A\arabic{equation}}

In order to show the invariance of the entanglement measure for even (odd) $%
n $ qubits, we need the following properties of the function $sgn$ ($%
sgn^{\ast }$). The functions $sgn$ and $sgn^{\ast }$ were recursively
defined in \cite{LDF07}. For readability, we redefine $sgn$ and $sgn^{\ast }$
as follows.

\textit{Definition of }$sgn$\textit{:}

\noindent\

\begin{equation}
sgn(n,i)=(-1)^{N(i)}\text{ when }0\leq i\leq 2^{n-3}-1.
\end{equation}

The definition of $N(i)$ is given in the last paragraph of this
introduction. Whereas, $N(i)$ is the number of the occurrences of
\textquotedblleft $1$\textquotedblright\ in the $n-$bit binary
representation $i_{n-1}...i_{1}i_{0}$ of $i$.

In fact, this definition of $sgn(n,i)$\ can be derived from the recursive
definition of $sgn(n,i)$ in \cite{LDF07} by using the following property 1
about $N(i)$.

\noindent \textit{Definition of }$sgn^{\ast }$:

\begin{equation*}
sgn^{\ast }(n,i)=%
\begin{cases}
(-1)^{N(i)} & \text{for }0\leq i\leq 2^{n-3}-1\text{,}\qquad \\
(-1)^{n+N(i)} & \text{for }2^{n-3}\leq i\leq 2^{n-2}-1\text{. }%
\end{cases}%
\end{equation*}

This definition of $sgn^{\ast }(n,i)$ can be derived from the recursive
definition of $sgn^{\ast }$ in \cite{LDF07}\ by using the following property
1 about $N(i)$.

It is straightforward to derive the following property 1 about $N(i)$ by
means of the definition of $N(i)$. The property 1 will be used in the proofs
of the following properties 2-5.

\noindent \textit{Property 1:}

\noindent (i). Assume that $0\leq k\leq 2^{n-l-2}-1$ and $0\leq j\leq
2^{l-2}-1$. Then$\ N(k+j\times 2^{n-l-1})=N(j)+N(k)$.

\noindent (ii). Assume that $0\leq k\leq 2^{n-l-2}-1$ and $0\leq t\leq
2^{l-3}-1$. Then $N(k+t2^{n-l})=N(k)+N(t)$ and $%
N(k+(2t+1)2^{n-l-1})=N(k)+N(t)+1$.

\noindent (iii). Assume that $0\leq k\leq 2^{n-l-2}-1$ and $0\leq j\leq
2^{l-2}-1$. Then, $N(2^{n-l-1}-1-k)=n-l-1-N(k)$ and $N((j+1)\times
2^{n-l-1}-1-k)=N(j)+n-l-1-N(k)$.

\begin{proof}
\noindent Proof of (i): \medskip \noindent

Let the binary number of $j$ be $j_{l-3}j_{l-4}...j_{1}j_{0}$, where $%
j_{i}\in \{0,1\}$. That is, $j=j_{l-3}\times 2^{l-3}+...+j_{1}\times
2^{1}+j_{0}\times 2^{0}$. $j\times 2^{n-l-1}=j_{l-3}\times
2^{n-4}+...+j_{1}\times 2^{n-l}+j_{0}\times 2^{n-l-1}$. Clearly, $%
N(j)=N(j\times 2^{n-l-1})$. \ Since $0\leq k\leq 2^{n-l-2}-1$, $N(k+j\times
2^{n-l-1})=N(j\times 2^{n-l-1})+N(k)=N(j)+N(k)$. \

\noindent Proof of (ii):

\medskip \noindent Let the binary representation of $t$ be $%
t_{l-4}...t_{1}t_{0}$ and the binary number of $k$ be $%
k_{n-l-3}...k_{1}k_{0} $, where $t_{i}$, $k_{i}\in \{0,1\}$. $%
k+(2t+1)2^{n-l-1}=k+t2^{n-l}+2^{n-l-1} $. The latter can be rewritten as $%
t_{l-4}2^{n-4}+...+t_{1}2^{n-l+1}+t_{0}2^{n-l}+2^{n-l-1}+k_{n-l-3}2^{n-l-3}+...+k_{0}2^{0}
$. It is obvious that $N(k+t2^{n-l}+2^{n-l-1})=N(k)+N(t)+1$. As well, $%
N(k+t2^{n-l})=N(k)+N(t)$.

\noindent Proof of (iii):

\medskip \noindent Let us calculate $N(2^{n-l-1}-1-k)$.\ The binary number
of $2^{n-l-1}-1$ is $\underbrace{1...1}_{n-l-1}$. That is, $%
2^{n-l-1}-1=2^{n-l-2}+...+2^{1}+2^{0}$. Let $k_{n-l-3}...k_{1}k_{0}$ be the
binary number of $k$, where $k_{i}\in \{0,1\}$. That is, $k=k_{n-l-3}\times
2^{n-l-3}+...+k_{1}\times 2^{1}+k_{0}\times 2^{0}$. Note that the binary
numbers of $2^{n-l-1}-1-k$\ and $k$\ are complementary. Hence, it is
straightforward that $N(2^{n-l-1}-1-k)=n-l-1-N(k)$.

$(j+1)\times 2^{n-l-1}-1-k=j\times 2^{n-l-1}+(2^{n-l-1}-1-k)$. Notice that $%
2^{n-l-2}\leq 2^{n-l-1}-1-k\leq 2^{n-l-1}-1$. It is intuitive that $%
N(j\times 2^{n-l-1}+(2^{n-l-1}-1-k))=N(j)+N(2^{n-l-1}-1-k)$. Therefore, $%
N((j+1)\times 2^{n-l-1}-1-k)=N(j)+n-l-1-N(k)$.
\end{proof}

The following properties 2-5 are used in proofs of Theorems 1 and 2. The
property 2 about $sgn$ follows the property 1 and the definition for $sgn$.

\noindent \textit{Property 2: }

Assume that $0\leq k\leq 2^{n-l-2}-1$ and $0\leq j\leq 2^{l-2}-1$. Then $%
sgn(n,(j+1)\times 2^{n-l-1}-1-k)=(-1)^{n+l+1}sgn(n,k+j\times 2^{n-l-1})$.

\begin{proof}
(1). Compute $sgn(n,k+j\times 2^{n-l-1})$. Since $k+j\times
2^{n-l-1}<2^{n-3}-1$, by the definition for $sgn$, $sgn(n,k+j\times
2^{n-l-1})=(-1)^{N(k+j\times 2^{n-l-1})}$. By (i) of property 1, $%
N(k+j\times 2^{n-l-1})=N(k)+N(j)$. Therefore $sgn(n,k+j\times
2^{n-l-1})=(-1)^{N(j)+N(k)}$. \

\noindent (2). Compute $sgn(n,(j+1)\times 2^{n-l-1}-1-k)$. Since $%
(j+1)\times 2^{n-l-1}-1-k\leq 2^{n-3}-1$, by the definition for $sgn$, $%
sgn(n,(j+1)\times 2^{n-l-1}-1-k)=(-1)^{N((j+1)\times 2^{n-l-1}-1-k)}$. By
(iii) of property 1, $sgn(n,(j+1)\times
2^{n-l-1}-1-k)=(-1)^{N(j)+n-l-1-N(k)} $.

Conclusively, $sgn(n,(j+1)\times 2^{n-l-1}-1-k)=(-1)^{n+l+1}sgn(n,k+j\times
2^{n-l-1})$.
\end{proof}

The property 3 about $sgn\ $and $sgn^{\ast }$ can be shown by means of the
property 1 and the definitions for $sgn$ and $sgn^{\ast }$.

\noindent \textit{Property 3:}

Assume that $0\leq k\leq 2^{n-l-2}-1$ and $0\leq t\leq 2^{l-3}-1$. Then

(i) $sgn(n,k+(2t+1)\times 2^{n-l-1})=-sgn(n,k+t\times 2^{n-l})$.

(ii) $sgn^{\ast }(n-1,k+(2t+1)\times 2^{n-l-1})=-sgn^{\ast }(n-1,k+t\times
2^{n-l})$.

\begin{proof}
\noindent Proof of (i): Since $k+(2t+1)\times 2^{n-l-1}\leq
2^{n-3}-2^{n-l-2}-1$, by the definition for $sgn$, $sgn(n,k+(2t+1)\times
2^{n-l-1})=(-1)^{N(k+(2t+1)\times 2^{n-l-1})}$. By (ii) of property 1, $%
sgn(n,k+(2t+1)\times 2^{n-l-1})=(-1)^{N(k)+N(t)+1}$. Similarly, $%
sgn(n,k+t\times 2^{n-l})=(-1)^{N(k)+N(t)}$.

\noindent Proof of (ii):

\begin{enumerate}
\item $0\leq t\leq 2^{l-4}-1$. Thus, $k+(2t+1)\times 2^{n-l-1}\leq 2^{n-4}-1$
and $k+t\times 2^{n-l}\leq 2^{n-4}-1$. By the definition for $sgn^{\ast }$
and (ii) of property 1, $sgn^{\ast }(n-1,k+(2t+1)\times
2^{n-l-1})=(-1)^{N(k)+N(t)+1}$ and

\begin{equation}
sgn^{\ast }(n-1,k+t\times 2^{n-l})=(-1)^{N(k)+N(t)}.  \label{property-5-1}
\end{equation}

\item $2^{l-4}\leq t\leq 2^{l-3}-1$. Thus, $2^{n-4}\leq k+t\times
2^{n-l}<2^{n-3}-1$ and $2^{n-4}<k+(2t+1)\times 2^{n-l-1}<2^{n-3}-1$. By the
definition for $sgn^{\ast }$ and (ii) of property 1, $sgn^{\ast
}(n-1,k+(2t+1)\times 2^{n-l-1})=(-1)^{n-1}(-1)^{N(k)+N(t)+1}$ and
\end{enumerate}

\begin{equation}
sgn^{\ast }(n-1,k+t\times 2^{n-l})=(-1)^{n-1}(-1)^{N(k)+N(t)}.
\label{property-5-2}
\end{equation}
\end{proof}

The property 4 about $sgn^{\ast }$ can be obtained from the property 1 and
the definition for $sgn^{\ast }$.

\noindent \textit{Property 4: }

Assume that $0\leq k\leq 2^{n-l-2}-1$ and $0\leq j\leq 2^{l-2}-1$. Then $%
sgn^{\ast }(n-1,(j+1)\times 2^{n-l-1}-1-k)=(-1)^{n+l+1}sgn^{\ast
}(n-1,k+j\times 2^{n-l-1})$.

\begin{proof}
\begin{enumerate}
\item $0\leq j\leq 2^{l-3}-1$. Since $(j+1)\times 2^{n-l-1}-1-k\leq
2^{n-4}-1 $, by the definition for $sgn^{\ast }$, $sgn^{\ast
}(n-1,(j+1)\times 2^{n-l-1}-1-k)=(-1)^{N((j+1)\times 2^{n-l-1}-1-k)}$. By
(iii) of property 1, $sgn^{\ast }(n-1,(j+1)\times
2^{n-l-1}-1-k)=(-1)^{N(j)+n-l-1-N(k)}$. As well, since $k+j\times
2^{n-l-1}<2^{n-4}-2^{n-l-2}-1$, by the definition for $sgn^{\ast }$, $%
sgn^{\ast }(n-1,k+j\times 2^{n-l-1})=(-1)^{N(k+j\times 2^{n-l-1})}$. By (i)
of property 1, $sgn^{\ast }(n-1,k+j\times 2^{n-l-1})=(-1)^{N(k)+N(j)}$.
Therefore, the property holds for this case.

\item $2^{l-3}\leq j\leq 2^{l-2}-1$. Thus, $2^{n-4}+2^{n-l-2}\leq
(j+1)\times 2^{n-l-1}-1-k\leq 2^{n-3}-1$ and $2^{n-4}+2^{n-l-2}-1\leq
k+j\times 2^{n-l-1}\leq 2^{n-3}-1$. By the definition for $sgn^{\ast }$ and
(iii) of property 1, $sgn^{\ast }(n-1,(j+1)\times
2^{n-l-1}-1-k)=(-1)^{n-1}(-1)^{N((j)+n-l-1-N(k)}$. By the definition for $%
sgn^{\ast }$ and (i) of property 1, $sgn^{\ast }(n-1,k+j\times
2^{n-l-1})=(-1)^{n-1}(-1)^{N(k)+N(j)}$. Therefore, this property holds for
this case.
\end{enumerate}
\end{proof}

It is not hard to derive the property 5 by means of the property 1 and the
definitions for $sgn$ and $sgn^{\ast }$.

\noindent \textit{Property 5: }

Assume that $0\leq k\leq 2^{n-l-2}-1$ and $0\leq t\leq 2^{l-3}-1$. When $n$
is odd and $l$ is odd or $n$ is even and $l$ is even,\ then the following
statements are true:

\noindent (i). $sgn^{\ast }(n-l,k)=(-1)^{N(k)}$,

\noindent (ii). $sgn(n,k+t\times 2^{n-l})=sgn^{\ast }(n-l,k)sgn(l,t)$,

\noindent (iii). $sgn^{\ast }(n-1,k+t\times 2^{n-l})=sgn^{\ast
}(n-l,k)sgn^{\ast }(l-1,t)$.

\begin{proof}
\noindent Proof of (i):

\begin{enumerate}
\item $0\leq k\leq 2^{n-l-3}-1$. By the definition for $sgn^{\ast }$, $%
sgn^{\ast }(n-l,k)=(-1)^{N(k)}$.

\item $2^{n-l-3}\leq k\leq 2^{n-l-2}-1$. By the definition for $sgn^{\ast }$%
, $sgn^{\ast }(n-l,k)=(-1)^{n-l}(-1)^{N(k)}$. When $n$ is odd and $l$ is odd
or $n$ is even and $l$ is even, clearly $sgn^{\ast }(n-l,k)=(-1)^{N(k)}$.
\end{enumerate}

\noindent From cases 1 and 2, this statement follows.

\noindent Proof of (ii):

\noindent Step 1. Compute $sgn(n,k+t\times 2^{n-l})$. Since $0\leq k+t\times
2^{n-l}\leq 2^{n-3}-2^{n-l}+2^{n-l-2}-1$, by the definition for $sgn$ and
(ii) of property 1, $sgn(n,k+t\times 2^{n-l})=(-1)^{N(k)+N(t)}$.

\noindent Step 2. Compute $sgn(l,t)$. By the definition for $sgn$, $%
sgn(l,t)=(-1)^{N(t)}$. From (i) of this property and steps 1 and 2, we can
conclude that (ii) holds.

\noindent Proof of (iii):

\begin{enumerate}
\item $0\leq t\leq 2^{l-4}-1$. By the definition for $sgn^{\ast }$, $%
sgn^{\ast }(l-1,t)=(-1)^{N(t)}$. By Eq. (\ref{property-5-1}), $sgn^{\ast
}(n-1,k+t\times 2^{n-l})=(-1)^{N(k)+N(t)}$. Therefore, by (i) of this
property, (iii) is true for this case.

\item $2^{l-4}\leq t\leq 2^{l-3}-1$. By the definition for $sgn^{\ast }$, $%
sgn^{\ast }(l-1,t)=(-1)^{l-1}(-1)^{N(t)}$. By Eq. (\ref{property-5-2}), $%
sgn^{\ast }(n-1,k+t\times 2^{n-l})=(-1)^{n-1}(-1)^{N(k)+N(t)}$. By (i) of
this property, it is not hard to see that (iii) holds for this case.
\end{enumerate}
\end{proof}

\section*{Appendix B. The proof of Theorem 1}

\setcounter{equation}{0} \renewcommand{\theequation}{B\arabic{equation}}

We show this theorem in three cases: case 1, $l=1$;\ case 2, $l=2$; case 3, $%
l\geq 3$.

Proof of $l=1$:

\begin{proof}
When $l=1$, $|\phi \rangle =b_{0}|0\rangle +b_{1}|1\rangle $. By solving $%
|\psi \rangle =|\phi \rangle _{1}\otimes |\omega \rangle _{2,...,n}$,\ we
obtain the following amplitudes:

\begin{equation}
a_{i}=b_{0}c_{i},\text{ }a_{2^{n-1}+i}=b_{1}c_{i},\text{ }0\leq i\leq
2^{n-1}-1.  \label{even-ampli-1}
\end{equation}

By substituting the amplitudes in Eq. (\ref{even-ampli-1}) into $\mathcal{I}%
^{\ast }(a,n)$ in Eq. (\ref{even-ver-2}),

$\mathcal{I}^{\ast }(a,n)=b_{0}b_{1}\sum_{i=0}^{2^{n-2}-1}sgn^{\ast
}(n,i)(c_{2i}c_{(2^{n-1}-1)-2i}-c_{2i+1}c_{(2^{n-1}-2)-2i})$

\ \ \ \ \ \ \ \ $=b_{0}b_{1}\sum_{i=0}^{2^{n-3}-1}sgn^{\ast
}(n,i)(c_{2i}c_{(2^{n-1}-1)-2i}-c_{2i+1}c_{(2^{n-1}-2)-2i})+$

\ \ \ \ \ \ \ \ \ \ \ $b_{0}b_{1}\sum_{i=2^{n-3}}^{2^{n-2}-1}sgn^{\ast
}(n,i)(c_{2i}c_{(2^{n-1}-1)-2i}-c_{2i+1}c_{(2^{n-1}-2)-2i})$.

Let $k=2^{n-2}-1-i$. Then the last sum can be rewritten as

$-b_{0}b_{1}\sum_{k=2^{n-3}-1}^{0}sgn^{\ast
}(n,2^{n-2}-1-k)(c_{2k}c_{(2^{n-1}-1)-2k}-c_{2k+1}c_{(2^{n-1}-2)-2k})$.

It is easy to demonstrate $sgn^{\ast }(n,2^{n-2}-1-k)=sgn^{\ast }(n,k)$ by
the definition of $sgn^{\ast }$. Thus, $\mathcal{I}^{\ast }(a,n)=0$ and $%
\tau (\psi )=0$.
\end{proof}

Proof of $l=2$:

\begin{proof}
In this case, $|\phi \rangle $ is a state of the first two qubits and $|\phi
\rangle $ $=$\ $\sum_{i=0}^{3}b_{i}|i\rangle $, $|\omega \rangle $ is a
state of the last $(n-2)$-qubits and $|\omega \rangle $ $=$\ $%
\sum_{i=0}^{2^{n-2}-1}c_{i}|i\rangle $. By the definition \cite{LDF07}, $%
\tau (\phi )=2\left\vert b_{0}b_{3}-b_{1}b_{2}\right\vert $, $\tau (\omega
)=2\left\vert \mathcal{I}^{\ast }(c,n-2)\right\vert $, and $\tau (\psi
)=2\left\vert \mathcal{I}^{\ast }(a,n)\right\vert $. We can write
\begin{equation}
|\psi \rangle =|\phi \rangle _{1,2}\otimes |\omega \rangle _{3,...,n}
\label{product-state}
\end{equation}%
By solving Eq. (\ref{product-state}), we obtain the following amplitudes:

\begin{equation}
a_{j}=b_{0}c_{j},\quad a_{2^{n-2}+j}=b_{1}c_{j},\quad
a_{2^{n-1}+j}=b_{2}c_{j},\quad a_{3\times 2^{n-2}+j}=b_{3}c_{j}
\label{amplitudes}
\end{equation}%
, where $0\leq j\leq 2^{n-2}-1$.

We rewrite $\mathcal{I}^{\ast }(a,n)=E_{1}+E_{2}$, where
\begin{equation}
E_{1}=\sum_{i=0}^{2^{n-3}-1}sgn^{\ast
}(n,i)(a_{2i}a_{(2^{n}-1)-2i}-a_{2i+1}a_{(2^{n}-2)-2i})
\end{equation}%
and
\begin{equation}
E_{2}=\sum_{i=2^{n-3}}^{2^{n-2}-1}sgn^{\ast
}(n,i)(a_{2i}a_{(2^{n}-1)-2i}-a_{2i+1}a_{(2^{n}-2)-2i}).
\end{equation}%
Let us compute $E_{1}$ as follows. Since $0\leq i\leq $ $2^{n-3}-1$, by Eq. (%
\ref{amplitudes})

\begin{eqnarray}
&&a_{2i}=b_{0}c_{2i},a_{(2^{n}-1)-2i}=b_{3}c_{(2^{n-2}-1)-2i},  \notag \\
&&a_{2i+1}=b_{0}c_{2i+1},a_{(2^{n}-2)-2i}=b_{3}c_{(2^{n-2}-2)-2i}.
\label{coefficients}
\end{eqnarray}%
By substituting the amplitudes in Eq. (\ref{coefficients}) into $E_{1}$, $%
E_{1}$ becomes

\begin{equation}
E_{1}=b_{0}b_{3}\sum_{i=0}^{2^{n-3}-1}sgn^{\ast
}(n,i)(c_{2i}c_{(2^{n-2}-1)-2i}-c_{2i+1}c_{(2^{n-2}-2)-2i})  \label{E1}
\end{equation}

In Eq. (\ref{E1}) let $E_{1}=E_{1}^{(1)}+E_{1}^{(2)}$, where
\begin{equation}
E_{1}^{(1)}=b_{0}b_{3}\sum_{i=0}^{2^{n-4}-1}sgn^{\ast
}(n,i)(c_{2i}c_{(2^{n-2}-1)-2i}-c_{2i+1}c_{(2^{n-2}-2)-2i})  \label{E11}
\end{equation}%
and
\begin{equation}
E_{1}^{(2)}=b_{0}b_{3}\sum_{i=2^{n-4}}^{2^{n-3}-1}sgn^{\ast
}(n,i)(c_{2i}c_{(2^{n-2}-1)-2i}-c_{2i+1}c_{(2^{n-2}-2)-2i}).
\end{equation}%
Let us demonstrate $E_{1}^{(2)}=E_{1}^{(1)}$. Let $k=(2^{n-3}-1)-i$. Then

\begin{equation}
E_{1}^{(2)}=-b_{0}b_{3}\sum_{k=2^{n-4}-1}^{0}sgn^{\ast
}(n,2^{n-3}-1-k)(c_{2k}c_{(2^{n-2}-1)-2k}-c_{2k+1}c_{(2^{n-2}-2)-2k}).
\label{E12}
\end{equation}%
When $0\leq k\leq $\ $2^{n-4}-1$, by the definition for $sgn^{\ast }$ and
(iii) of property 1 in Appendix A, then $sgn^{\ast
}(n,2^{n-3}-1-k)=-sgn^{\ast }(n,k)$. Thus, $E_{1}^{(2)}=E_{1}^{(1)}$ and $%
E_{1}=2E_{1}^{(1)}$.\

Next we show $E_{1}=2b_{0}b_{3}$ $\mathcal{I}^{\ast }(c,n-2)$. For this
purpose, we only need to show $sgn^{\ast }(n,i)=sgn^{\ast }(n-2,i)$ provided
that $0\leq i\leq 2^{n-4}-1$. The definition for $sgn^{\ast }$ in Appendix A
asserts this.

Similarly, we can derive $E_{2}=-2b_{1}b_{2}$ $\mathcal{I}^{\ast }(c,n-2)$.
Thus, $\mathcal{I}^{\ast }(a,n)=2(b_{0}b_{3}-b_{1}b_{2})$ $\mathcal{I}^{\ast
}(c,n-2)$. Conclusively, $\tau (\psi )=\tau (\phi )\tau (\omega )$.
\end{proof}

Proof for $l\geq 3$:

\begin{proof}
We write

\begin{equation}
|\psi \rangle =|\phi \rangle _{1,...,l}\otimes |\omega \rangle
_{(l+1),...,n}.  \label{g-product-even}
\end{equation}

By solving equation Eq. (\ref{g-product-even}), we obtain the following
amplitudes:

\begin{equation}
\quad a_{k\times
2^{n-l}+i}=b_{k}c_{i},k=0,1,...,(2^{l}-1),i=0,1,...,(2^{n-l}-1).
\label{g-amplitudes}
\end{equation}

We rewrite $\mathcal{I}^{\ast }(a,n)$ as $\mathcal{I}^{\ast }(a,n)=$ $%
\sum_{j=0}^{2^{l-1}-1}\Delta _{j},$ where

\begin{eqnarray}
\Delta _{j} &=&\sum_{i=j\times 2^{n-l-2}}^{(j+1)\times
2^{n-l-2}-1}sgn(n,i)[(a_{2i}a_{(2^{n}-1)-2i}-a_{2i+1}a_{(2^{n}-2)-2i})
\notag \\
&&+(a_{(2^{n-1}-2)-2i}a_{(2^{n-1}+1)+2i}-a_{(2^{n-1}-1)-2i}a_{2^{n-1}+2i})].
\end{eqnarray}

By substituting the amplitudes in Eq. (\ref{g-amplitudes}) into $\Delta
_{2j} $ and $\Delta _{2j+1}$, we get

\begin{eqnarray}
\Delta _{2j} &=&\sum_{k=0}^{2^{n-l-2}-1}sgn(n,k+j\times 2^{n-l-1})\times
\notag \\
&&(b_{j}b_{2^{l}-1-j}-b_{2^{l-1}+j}b_{2^{l-1}-1-j})\times  \notag \\
&&[(c_{2k}c_{(2^{n-l}-1)-2k}-c_{2k+1}c_{(2^{n-l}-2)-2k})],
\end{eqnarray}

and

\begin{eqnarray}
\Delta _{2j+1} &=&-\sum_{k=2^{n-l-2}-1}^{0}sgn(n,(j+1)\times
2^{n-l-1}-1-k)\times  \notag \\
&&(b_{j}b_{2^{l}-1-j}-b_{2^{l-1}+j}b_{2^{l-1}-1-j})\times  \notag \\
&&[(c_{2k}c_{(2^{n-l}-1)-2k}-c_{2k+1}c_{(2^{n-l}-2)-2k})].
\end{eqnarray}

When $l$ is odd,\ by property 2 in Appendix A, then $\Delta _{2j+1}=-\Delta
_{2j}$, $j=0$, $1$, $...$, $2^{l-2}-1$. Hence, $\mathcal{I}^{\ast }(a,n)=0$.
Thus, $\tau (\psi )=0$. When $l$ is even, by property 2 in Appendix A, then $%
\Delta _{2j+1}=\Delta _{2j}$, $j=0$, $1$, $...$, $2^{l-2}-1$. Therefore

\begin{equation}
\mathcal{I}^{\ast }(a,n)=2\sum_{j=0}^{2^{l-2}-1}\Delta
_{2j}=2\sum_{t=0}^{2^{l-3}-1}(\Delta _{4t}+\Delta _{4t+2}).  \label{even-n1}
\end{equation}

By (i) of property 3 in Appendix A, from Eq. (\ref{even-n1}),
\begin{eqnarray}
\mathcal{I}^{\ast }(a,n)=2
&&\sum_{t=0}^{2^{l-3}-1}\{[(b_{2t}b_{2^{l}-1-2t}-b_{2t+1}b_{2^{l}-2-2t})+
\notag \\
&&(b_{2^{l-1}-2-2t}b_{2^{l-1}+1+2t}-b_{2^{l-1}-1-2t}b_{2^{l-1}+2t})]\times
\notag \\
&&\sum_{k=0}^{2^{n-l-2}-1}sgn(n,k+t\times 2^{n-l})\times  \notag \\
&&[c_{2k}c_{(2^{n-l}-1)-2k}-c_{2k+1}c_{(2^{n-l}-2)-2k}]\}.  \label{even-n2}
\end{eqnarray}

By (ii) of property 5 in Appendix A, from Eq. (\ref{even-n2}), we obtain
\begin{equation}
\mathcal{I}^{\ast }(a,n)=2\mathcal{I}^{\ast }(b,l)\mathcal{I}^{\ast }(c,n-l).
\end{equation}

Therefore, $\tau (\psi )=\tau (\phi )\tau (\omega )$.
\end{proof}

\section*{Appendix C. The proof of theorem 2}

\setcounter{equation}{0} \renewcommand{\theequation}{C\arabic{equation}}

When $l=1$, see \cite{LDF07}. When $l=2$, the proof is omitted. \ Next let
us consider that $l\geq 3$.

\begin{proof}
Step 1. Compute $\overline{\mathcal{I}}(a,n)$.

We rewrite $\overline{\mathcal{I}}(a,n)$ in Eq. (\ref{odd-iv}) as $\overline{%
\mathcal{I}}(a,n)=\sum_{j=0}^{2^{l-1}-1}\Omega _{j},$ where

\begin{eqnarray}
\Omega _{j} &=&\sum_{i=j\times 2^{n-l-2}}^{(j+1)\times
2^{n-l-2}-1}sgn(n,i)[(a_{2i}a_{(2^{n}-1)-2i}-a_{2i+1}a_{(2^{n}-2)-2i})
\notag \\
&&-(a_{(2^{n-1}-2)-2i}a_{(2^{n-1}+1)+2i}-a_{(2^{n-1}-1)-2i}a_{2^{n-1}+2i})].
\end{eqnarray}

By substituting the amplitudes in Eq. (\ref{g-amplitudes}) into $\Omega
_{2j} $ and $\Omega _{2j+1}$, we obtain

\begin{eqnarray}
\Omega _{2j} &=&\sum_{k=0}^{2^{n-l-2}-1}sgn(n,k+j\times 2^{n-l-1})\times
\notag \\
&&(b_{j}b_{2^{l}-1-j}+b_{2^{l-1}+j}b_{2^{l-1}-1-j})\times  \notag \\
&&[(c_{2k}c_{(2^{n-l}-1)-2k}-c_{2k+1}c_{(2^{n-l}-2)-2k})],
\end{eqnarray}

and

\begin{eqnarray}
\Omega _{2j+1} &=&-\sum_{k=2^{n-l-2}-1}^{0}sgn(n,(j+1)\times
2^{n-l-1}-1-k)\times  \notag \\
&&(b_{j}b_{2^{l}-1-j}+b_{2^{l-1}+j}b_{2^{l-1}-1-j})\times  \notag \\
&&[(c_{2k}c_{(2^{n-l}-1)-2k}-c_{2k+1}c_{(2^{n-l}-2)-2k})].
\end{eqnarray}

When $l$ is even,\ by property 2 in Appendix A, then $\Omega _{2j+1}=-\Omega
_{2j}$, $j=0$, $1$, $...$, $2^{l-2}-1$. Hence, $\overline{\mathcal{I}}%
(a,n)=0 $. When $l$ is odd, by property 2 in Appendix A, then $\Omega
_{2j+1}=\Omega _{2j}$, $j=0$, $1$, $...$, $2^{l-2}-1$. Therefore

\begin{equation}
\overline{\mathcal{I}}(a,n)=2\sum_{j=0}^{2^{l-2}-1}\Omega
_{2j}=2\sum_{t=0}^{2^{l-3}-1}(\Omega _{4t}+\Omega _{4t+2}).  \label{step11}
\end{equation}

By (i) of property 3 in Appendix A, from Eq. (\ref{step11}), we obtain
\begin{eqnarray}
\overline{\mathcal{I}}(a,n)=2
&&\sum_{t=0}^{2^{l-3}-1}\{[(b_{2t}b_{2^{l}-1-2t}-b_{2t+1}b_{2^{l}-2-2t})-
\notag \\
&&(b_{2^{l-1}-2-2t}b_{2^{l-1}+1+2t}-b_{2^{l-1}-1-2t}b_{2^{l-1}+2t})]\times
\notag \\
&&\sum_{k=0}^{2^{n-l-2}-1}sgn(n,k+t\times 2^{n-l})\times  \notag \\
&&[c_{2k}c_{(2^{n-l}-1)-2k}-c_{2k+1}c_{(2^{n-l}-2)-2k}]\}.  \label{step12}
\end{eqnarray}

By (ii) of property 5 in Appendix A, from Eq. (\ref{step12}), we obtain
\begin{equation}
\overline{\mathcal{I}}(a,n)=2\overline{\mathcal{I}}(b,l)\mathcal{I}^{\ast
}(c,n-l).  \label{step13}
\end{equation}

Step 2. Compute $\mathcal{I}_{+2^{n-1}}^{\ast }(a,n-1)$.

We can rewrite $\mathcal{I}_{+2^{n-1}}^{\ast }(a,n-1)$ as $\mathcal{I}%
_{+2^{n-1}}^{\ast }(a,n-1)=\sum_{j=0}^{2^{l-1}-1}Q_{j},$ where

\begin{equation}
Q_{j}=\sum_{i=j\times 2^{n-l-2}}^{(j+1)\times 2^{n-l-2}-1}sgn^{\ast
}(n-1,i)(a_{2^{n-1}+2i}a_{(2^{n}-1)-2i}-a_{2^{n-1}+1+2i}a_{(2^{n}-2)-2i}).
\end{equation}

By substituting the amplitudes in Eq. (\ref{g-amplitudes}) into $Q_{2j}$ and
$Q_{2j+1}$, we get

\begin{eqnarray}
Q_{2j} &=&\sum_{k=0}^{2^{n-l-2}-1}sgn^{\ast }(n-1,k+j\times 2^{n-l-1})\times
\notag \\
&&(b_{2^{l-1}+j}b_{2^{l-1}-1-j})[(c_{2k}c_{(2^{n-l}-1)-2k}-c_{2k+1}c_{(2^{n-l}-2)-2k})],
\end{eqnarray}

and

\begin{eqnarray}
&&Q_{2j+1} =-\sum_{k=2^{n-l-2}-1}^{0}sgn^{\ast }(n-1,(j+1)
2^{n-l-1}-1-k)\times  \notag \\
&&(b_{2^{l-1}+j}b_{2^{l-1}-1-j})[(c_{2k}c_{(2^{n-l}-1)-2k}-c_{2k+1}c_{(2^{n-l}-2)-2k})].
\end{eqnarray}

When $l$\ is even, by property 4 in Appendix A, then $Q_{2j+1}=-Q_{2j}$, $%
j=0 $, $1$, $...$, $2^{l-2}-1$. Hence, $\mathcal{I}_{+2^{n-1}}^{\ast
}(a,n-1)=0$. When $l$ is odd, by property 4 in Appendix A, then $%
Q_{2j+1}=Q_{2j}$, $j=0$, $1$, $...$, $2^{l-2}-1$. Therefore

\begin{equation}
\mathcal{I}_{+2^{n-1}}^{\ast
}(a,n-1)=2\sum_{j=0}^{2^{l-2}-1}Q_{2j}=2%
\sum_{t=0}^{2^{l-3}-1}(Q_{4t}+Q_{4t+2}).  \label{step21}
\end{equation}

By (ii) of property 3 in Appendix A, from Eq. (\ref{step21}), we obtain

\begin{eqnarray}
&&\mathcal{I}_{+2^{n-1}}^{\ast }(a,n-1)=  \notag \\
&&2%
\sum_{t=0}^{2^{l-3}-1}[(b_{2^{l-1}+2t}b_{2^{l-1}-1-2t}-b_{2^{l-1}+1+2t}b_{2^{l-1}-2-2t})\times
\notag \\
&&\sum_{k=0}^{2^{n-l-2}-1}sgn^{\ast }(n-1,k+t\times
2^{n-l})(c_{2k}c_{2^{n-l}-1-2k}-c_{2k+1}c_{2^{n-l}-2-2k})].  \label{step22}
\end{eqnarray}

By (iii) of property 5 in Appendix A, from Eq. (\ref{step22}), we get
\begin{equation}
\mathcal{I}_{+2^{n-1}}^{\ast }(a,n-1)=2\mathcal{I}_{+2^{n-1}}^{\ast }(b,l-1)%
\mathcal{I}^{\ast }(c,n-l).  \label{step23}
\end{equation}

Step 3. Compute $\mathcal{I}^{\ast }(a,n-1)$.

We rewrite $\mathcal{I}^{\ast }(a,n-1)$ as $\mathcal{I}^{\ast
}(a,n-1)=\sum_{j=0}^{2^{l-1}-1}R_{j}$, where

\begin{equation}
R_{j}=\sum_{i=j\times 2^{n-l-2}}^{(j+1)\times 2^{n-l-2}-1}sgn^{\ast
}(n-1,i)(a_{2i}a_{(2^{n-1}-1)-2i}-a_{2i+1}a_{(2^{n-1}-2)-2i}).
\end{equation}

By substituting the amplitudes in Eq. (\ref{g-amplitudes}) into $R_{2j}$ and
$R_{2j+1}$, we get

\begin{eqnarray}
R_{2j} &=&\sum_{k=0}^{2^{n-l-2}-1}sgn^{\ast }(n-1,k+j\times 2^{n-l-1})\times
\notag \\
&&(b_{j}b_{2^{l-1}-1-j})[(c_{2k}c_{(2^{n-l}-1)-2k}-c_{2k+1}c_{(2^{n-l}-2)-2k})],
\end{eqnarray}

and

\begin{eqnarray}
R_{2j+1} &=&-\sum_{k=2^{n-l-2}-1}^{0}sgn^{\ast }(n-1,(j+1)\times
2^{n-l-1}-1-k)\times  \notag \\
&&(b_{j}b_{2^{l-1}-1-j})[(c_{2k}c_{(2^{n-l}-1)-2k}-c_{2k+1}c_{(2^{n-l}-2)-2k})].
\end{eqnarray}

When $l$\ is even, by property 4 in Appendix A, then $R_{2j+1}=-R_{2j}$, $%
j=0 $, $1$, $...$, $2^{l-2}-1$. Hence, $\mathcal{I}^{\ast }(a,n-1)=0$. When $%
l$ is odd, by property 4 in Appendix A, then $R_{2j+1}=R_{2j}$, $j=0$, $1$, $%
... $, $2^{l-2}-1$. Therefore

\begin{equation}
\mathcal{I}^{\ast
}(a,n-1)=2\sum_{j=0}^{2^{l-2}-1}R_{2j}=2%
\sum_{t=0}^{2^{l-3}-1}(R_{4t}+R_{4t+2}).  \label{step31}
\end{equation}

By (ii) of property 3 in Appendix A, from Eq. (\ref{step31}), we get

\begin{eqnarray}
&&\mathcal{I}^{\ast }(a,n-1)=  \notag \\
&&2\sum_{t=0}^{2^{l-3}-1}[(b_{2t}b_{2^{l-1}-1-2t}-b_{2t+1}b_{2^{l-1}-2-2t})
\times  \notag \\
&&\sum_{k=0}^{2^{n-l-2}-1}sgn^{\ast }(n-1,k+t 2^{n-l})\times  \notag \\
&&(c_{2k}c_{2^{n-l}-1-2k}-c_{2k+1}c_{2^{n-l}-2-2k})].  \label{step32}
\end{eqnarray}

By (iii) of property 5 in Appendix A, from Eq. (\ref{step32}), we get
\begin{equation}
\mathcal{I}^{\ast }(a,n-1)=2\mathcal{I}^{\ast }(b,l-1)\mathcal{I}^{\ast
}(c,n-l).  \label{step33}
\end{equation}

From steps 1, 2 and 3, it is obvious that by the definition of $\tau (\psi )$%
, $\tau (\psi )=0$ whenever $l$\ is even. While $l$\ is odd, by substituting
Eqs. (\ref{step13}), (\ref{step23}) and (\ref{step33}) into $\tau (\psi )$
in Eq. (\ref{odd-residual-def}),
\begin{eqnarray}
&&\tau (\psi )=16|(\overline{\mathcal{I}}(b,l))^{2}-4\mathcal{I}^{\ast
}(b,l-1)\mathcal{I}_{+2^{l-1}}^{\ast }(b,l-1)|\times  \notag \\
&&\left\vert \mathcal{I}^{\ast }(c,n-l)\right\vert ^{2}=\tau (\phi )\tau
^{2}(\omega ).
\end{eqnarray}
\end{proof}

\end{document}